\documentclass[twocolumn,superscriptaddress,aps,preprintnumbers,amsmath,amssymb,nofootinbib,a4paper
]{revtex4-2}

\usepackage{amsfonts,amsmath,amssymb,ascmac,bm,tensor}

\usepackage{fnpct} 
\usepackage{comment}
\usepackage{ifpdf}
\usepackage{slashed}
\usepackage{color}
\usepackage[mathscr]{eucal}
\usepackage{fourier}

\usepackage[utf8]{inputenc}
\usepackage{physics}
\usepackage{cancel}
\usepackage{subfig}
\usepackage{wasysym}

\usepackage{lipsum}

\ifpdf
  \usepackage{graphicx}   
  \usepackage[bookmarksopen,colorlinks=true,linkcolor=bblue,citecolor=bblue,urlcolor=ppink]{hyperref}

\else   
\fi

\usepackage{cancel}
\usepackage{ulem}
\usepackage{amsfonts}
\usepackage{amssymb}
\usepackage{graphicx}
\usepackage{amsmath,bm}
\usepackage{enumerate}
\usepackage{mathtools}
\usepackage{tikz} \usetikzlibrary{calc}
\usepackage{subfloat}
\usepackage{subfig}
\usepackage{xcolor}
\usepackage{float}
\usepackage{color}
\usepackage{here}
\usepackage{booktabs}
\usepackage{siunitx}
\usepackage[table]{xcolor}
\makeatletter
\renewcommand\@biblabel[1]{\textcolor{red}{[#1]}}

\let\old@bibitem\@bibitem
\renewcommand{\@bibitem}[2][]{%
  \old@bibitem[#1]{#2}%
  \color{blue}
}
\makeatother

\usepackage{adjustbox}
\usepackage{graphicx,tikz}
\usetikzlibrary{mindmap}
\usepackage{mathrsfs}
\usepackage{float,epsfig}
\usepackage{dcolumn}
\usepackage{pgfplots}
\usepackage{graphicx}
\usepackage{bm}
\usepackage{amsthm}
\usetikzlibrary{shadows}

\newcommand*\eye{%
       \scalebox{0.35}{
    \tikzset{every picture/.style={line width=0.75pt}} 
    \begin{tikzpicture}[x=0.75pt,y=0.75pt,yscale=-1,xscale=1]
    \draw  [line width=1.5]  (300,100.33) .. controls (326,122) and (352,135) .. (378,139.33) .. controls (352,143.67) and (326,156.67) .. (300,178.33) ;
    \draw  [fill={rgb, 255:red, 0; green, 0; blue, 0 }  ,fill opacity=1 ] (308.94,116.33) .. controls (313.87,116.33) and (317.86,125.51) .. (317.85,136.83) .. controls (317.84,148.15) and (313.84,157.33) .. (308.91,157.33) .. controls (303.99,157.32) and (300,148.14) .. (300.01,136.82) .. controls (300.02,125.5) and (304.02,116.32) .. (308.94,116.33) -- cycle ;
    \draw  [draw opacity=0][line width=1.5]  (314.84,166.6) .. controls (311.87,164.64) and (309.14,162.18) .. (306.76,159.24) .. controls (295.12,144.82) and (296.6,124.33) .. (310.07,113.45) .. controls (311.48,112.32) and (312.96,111.33) .. (314.5,110.49) -- (331.14,139.55) -- cycle ; \draw  [line width=1.5]  (314.84,166.6) .. controls (311.87,164.64) and (309.14,162.18) .. (306.76,159.24) .. controls (295.12,144.82) and (296.6,124.33) .. (310.07,113.45) .. controls (311.48,112.32) and (312.96,111.33) .. (314.5,110.49) ;
    \draw  [fill={rgb, 255:red, 255; green, 255; blue, 255 }  ,fill opacity=1 ] (304.43,124.2) .. controls (306.09,124.25) and (307.32,128.01) .. (307.18,132.6) .. controls (307.05,137.19) and (305.59,140.88) .. (303.93,140.83) .. controls (302.27,140.78) and (301.03,137.02) .. (301.17,132.43) .. controls (301.31,127.83) and (302.76,124.15) .. (304.43,124.2) -- cycle ;
    \end{tikzpicture}
    }\,}

\definecolor{red}{rgb}{1,0,0}
\definecolor{darkred}{rgb}{0.6,0,0}
\definecolor{darkgreen}{rgb}{0.992447,0.623778,0.034597}
\definecolor{ppink}{rgb}{1,0.4,0.4}
\definecolor{bblue}{rgb}{0.284602,0.317763,0.963947}
\definecolor{brown}{rgb}{0.5 ,0, 0.7}
\definecolor{blue2}{rgb}{0.2 ,0.2, 0.85}

\makeatletter
\newcommand\footnoteref[1]{\protected@xdef\@thefnmark{\ref{#1}}\@footnotemark}
\makeatother
\allowdisplaybreaks[1]

\definecolor{lime}{HTML}{A6CE39}
\newcommand{\orcidicon}{%
	\begin{tikzpicture}
	\draw[lime, fill=lime] (0,0)
	circle [radius=0.16]
	node[white] {{\fontfamily{qag}\selectfont \tiny ID}};
	\draw[white, fill=white] (-0.0625,0.095)
	circle [radius=0.007];
	\end{tikzpicture}   \hspace{-2mm}
}
\newcommand\orcidHasan{{\href{https://orcid.org/0000-0001-7408-0910}{\orcidicon}}}
\newcommand\orcidKarima{{\href{https://orcid.org/0000-0001-5419-8516}{\orcidicon}}}
\newcommand\orcidFaical{{\href{https://orcid.org/0000-0002-2977-0821}{\orcidicon}}}
\begin{document}

\title{
Modified Unruh Thermodynamics in Emergent Gravity: Finite Heat Capacity and Rényi Entropy
    }

\author{F. Barzi\orcidFaical\!\!}
\email{faical.barzi@edu.uiz.ac.ma}
\affiliation{\small LPTHE, Physics Department, Faculty of Sciences,  Ibnou Zohr University, Agadir, Morocco.}
\affiliation{\small CRMEF, Regional Center for Education and Training Professions, Marrakesh, Morocco. }
\author{H. El Moumni \orcidHasan}	
\email{h.elmoumni@uiz.ac.ma (Corresponding author)}
\author{K.  Masmar \orcidKarima}
\email{karima.masmar@gmail.com}
\affiliation{\small LPTHE, Physics Department, Faculty of Sciences,  Ibnou Zohr University, Agadir, Morocco.}

\begin{abstract}

We show that Jacobson’s thermodynamic derivation of Einstein’s equations remains valid when local Rindler horizons are modeled as finite heat-capacity systems, resolving the infinite-bath assumption of Unruh thermodynamics. The horizon entropy then takes the form of Rényi entropy with nonextensivity parameter $\lambda\sim C^{-1}$, or equivalently a new “Einstein entropy” that uniquely preserves Einstein’s equations for arbitrary $C$. In both cases the Unruh temperature is modified to
\begin{equation*}
T_\text{mod}=\frac{\hbar\kappa}{2\pi}\left(1+\frac{S}{C}\right),
\end{equation*}
establishing a universal link between finite-capacity thermodynamics and generalized entropies. We further derive a corrected scalar Einstein equation with an upper bound on horizon energy flux, suggesting testable signatures in heavy-ion collisions, spin-polarization experiments, and analog gravity.


\end{abstract}

\date{\today}
\maketitle

\section{Introduction}

The Unruh effect \cite{unruh1976} establishes that a uniformly accelerated observer perceives the Minkowski vacuum as a thermal bath at temperature $T_U=\hbar a / 2\pi k_B c$. This cornerstone result, however, relies on the idealization of an infinite heat capacity reservoir, where energy exchange leaves the temperature unchanged. Realistic physical systems—from diamond-shaped causal domains to quark–gluon plasmas \cite{Teslyk2018}—instead exhibit a finite heat capacity $C$, requiring modifications to the standard thermodynamic picture. Such corrections have direct implications for Jacobson’s emergent-gravity program, where Einstein’s equations arise from the Clausius relation applied to local Rindler horizons \cite{jacobson1995}.

Here we demonstrate that accounting for finite heat capacity preserves the thermodynamic emergence of Einstein’s equations while revealing a natural connection to generalized entropies. The Clausius relation acquires the form $\delta Q = T_R\, dS_R$, with $S_R$ the Rényi entropy and $T_R$ the corresponding temperature. Equivalently, we introduce an “Einstein entropy” that depends explicitly on the horizon’s heat capacity and reproduces Einstein’s equations exactly for all $C$. In both cases, the Unruh temperature is modified, establishing gravity as an emergent phenomenon of nonextensive spacetime thermodynamics. Independent motivations for this framework arise in several contexts. In conformal field theory, finite-temperature Rindler space exhibits acceleration-enhanced thermality and instabilities near $T=T_U$ \cite{Diakonov2025}. In quantum gravity, the Bekenstein bound $S \leq 2\pi k_B E R / \hbar c$ implies $C \sim A$ for horizons, in contradiction with the infinite-bath assumption. Empirical hints also appear: heavy-ion collisions at RHIC and LHC reach accelerations $a \sim 10^{32}\,\text{m/s}^2(\sim1\,GeV)$ with maximum effective Unruh temperatures around $10^{12}\,K(\sim200\,\text{MeV})$ and discontinuous heat capacities \cite{Teslyk2018,Prokhorov2025}; spin polarization in storage rings and cold-atom experiments may also be used to probe similar finite-$C$ signatures\cite{Graham2021,Lous2022}. Finally, in holography, finite-capacity corrections can be implemented through finite-dimensional Hilbert spaces \cite{akers2019holographic}.

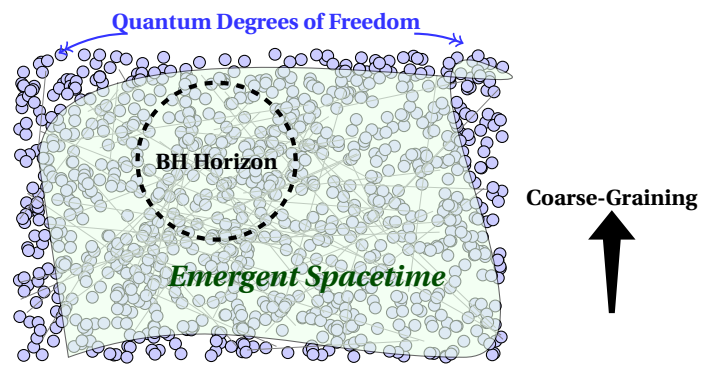
\begin{figure}[!htb]
\vspace{-0.6cm}
\centering
\hspace{0.5cm}
\begin{tikzpicture}[scale=1.3]
  \foreach \i in {1,2,...,500} {
    \draw[fill=blue!20] (rnd*4.9, rnd*3.1) circle (0.06);
  }
  \node[blue!80] at (2.5,3.4) {\textbf{Quantum Degrees of Freedom}};
  \draw[blue!80, thick, ->>] (1.1,3.3) parabola (0.4,3.15);
  \draw[blue!80, thick, ->>] (4,3.3) parabola (4.5,3.2);
  \draw[fill=green!8, opacity=0.6] (0.5,0.) to[out=20,in=180] (4.5,0.) 
    to[out=0,in=160] (4.7,3) to[out=-20,in=90] (0.2,2) -- cycle;
  \node[green!30!black,] at (2.9,.8) {\textbf{\large\textit{Emergent Spacetime}}};
  \draw[black, dashed,ultra thick] (2,2) circle (0.8);
  \node at (2,2) {\textbf{BH Horizon}};
  \fill[black,scale=0.3] (20,5) -- 
                      (19.2,4) -- 
                      (19.7,4)--
                      (19.9,1.5) -- 
                      (20,1.5) -- 
                      (20.1,1.5) -- 
                      (20.3,4) -- 
                      (20.8,4)--
                 cycle;
 
  \node[black] at (6,1.6) {\textbf{Coarse-Graining}};
\end{tikzpicture}
\caption{\it \small Spacetime geometry emerging from entangled quantum degrees of freedom. Black hole horizons encode entanglement entropy $S = A/4G\hbar$.}
\label{fig:emergence}
\end{figure}

In a seminal work, Jacobson showed that Einstein’s equations can be derived as an equation of state by applying the Clausius relation $\delta Q = T\,dS$ to local Rindler horizons \cite{jacobson1995}. This perspective elevates gravity from a fundamental interaction to an emergent phenomenon, Fig.\ref{fig:emergence}, with spacetime geometry adjusting to maintain thermodynamic equilibrium. In this view, gravitons play the role of phonons-collective excitations of underlying microscopic degrees of freedom-rather than fundamental particles. The derivation relies on the equivalence principle, which identifies local acceleration with a uniform gravitational field, and on horizon thermodynamics encoded by the Unruh effect and the Bekenstein–Hawking entropy.

The derivation hinges on three thermodynamic inputs:
\begin{align}
T_U &= \frac{a}{2\pi} \quad \text{(Unruh effect)}, \label{eq:unruh}\\
S &= \eta A \quad \text{(Bekenstein–Hawking entropy)}, \label{eq:entropy}\\
\delta Q &= T\, dS \quad \text{(Clausius relation)}, \label{eq:clausius}
\end{align}
with $\eta=(4\hbar G)^{-1}$. Requiring Eqs.\eqref{eq:unruh}–\eqref{eq:clausius} to hold for all local Rindler horizons leads directly to Einstein’s field equations as an equation of state \cite{jacobson1995}. Crucially, this argument assumes that the accelerated observer interacts with an infinite-capacity heat bath, so that horizon energy exchange does not affect the Unruh temperature. In realistic settings, however, horizons possess only a finite number of degrees of freedom, with heat capacity scaling as $C \sim A$ rather than $C \to \infty$. Finite-size effects manifest in bounded causal diamonds, where observers perceive a modified temperature. Such finite-$C$ corrections imply discrete energy exchange with the horizon and motivate a reformulation of the Clausius relation. We show that the appropriate replacement is
\begin{equation}
    \delta Q= T_R dS_R
\end{equation}
where $S_R$ is the Rényi entropy and $T_R$ the corresponding temperature, thereby preserving the thermodynamic emergence of gravity in the finite-capacity regime.

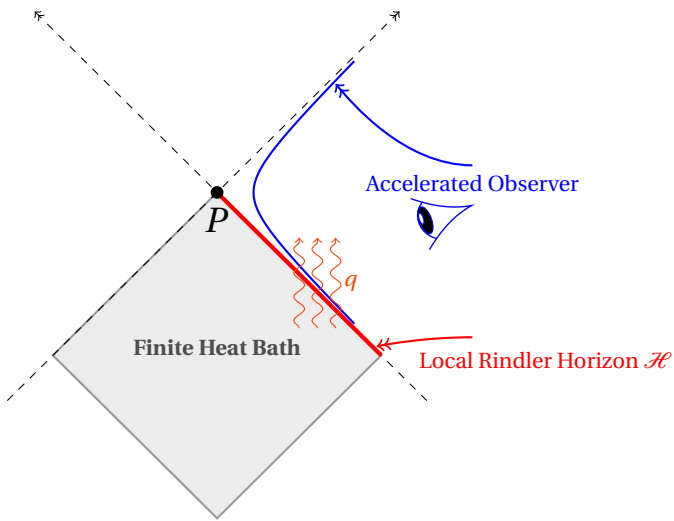
\begin{figure}[!htb]
\centering
\hspace{2.4cm}
\begin{tikzpicture}[scale=1.1]
\tikzset{every arrow/.style={line width=2pt, scale=5}}
   \fill[gray!15] (0,0) -- (1.8,-1.8) -- (0,-3.6) -- (-1.8,-1.8) -- cycle;
  \draw[gray!80,thick] (1.8,-1.8) -- (0,-3.6) -- (-1.8,-1.8) --(0,0)-- cycle;
 
  \draw[blue,  thick, domain=-2:2., smooth, variable=\t] 
        plot ({0.4*cosh(1*\t)}, {0.4*sinh(1*\t)}) node[above right]{};
 \node[black!70] at (0,-1.7){\textbf{Finite Heat Bath}};
  \draw[dashed, ->>] (-2.3,-2.3) -- (0,0) -- (2,2);
  \draw[dashed, ->>] (2.3,-2.3) -- (0,0) -- (-2,2);

  \draw[red, thick, ->>] (2.8,-1.6) parabola (1.75,-1.7);
  \node[red] at (3.6,-1.85) {Local Rindler Horizon $\mathcal{H}$};
   \draw[red, ultra thick] (1.8,-1.8) -- (0,0);
  
  \draw[[->, orange!50!red,decorate, decoration={snake, amplitude=2pt, segment length=9pt}] (1.1,-1.5) -- (1.1,-0.5) node[midway,right] {};
  \draw[[->, orange!50!red,decorate, decoration={snake, amplitude=2pt, segment length=9pt}] (0.9,-1.5) -- (0.9,-0.5) node[midway,right]{} ;
  \draw[[->, orange!50!red,decorate, decoration={snake, amplitude=2pt, segment length=9pt}] (1.3,-1.5) -- (1.3,-0.5) node[midway,right] {$q$};

  \draw[blue, thick, ->>] (2.8,0.3) parabola (1.3,1.2);
  \node[blue] at (2.8,0.1) {Accelerated Observer};
\node[blue, rotate=20] at (2.5,-0.25) {\eye };
 
\fill (0,0) circle (2pt);
  \node[black, font=\Large] at (0,-0.3){$P$};
\end{tikzpicture}
\caption{\it \small Causal structure in the neighborhood of spacetime point $P$ and the local Rindler horizon through $P$. The accelerated observer perceives $\mathcal{H}$ as a causal boundary with Unruh temperature $T_U$.}
\label{fig:horizon}
\end{figure}

Local Rindler horizons provide the setting for Jacobson’s thermodynamic derivation. By the equivalence principle, spacetime near any point $P$ can be approximated as flat, allowing the construction of a local causal horizon $\mathcal{H}$, Fig.\ref{fig:horizon}. The key thermodynamic ingredients are the Unruh temperature, $T_U=\hbar a/2\pi$ with $a$ the proper acceleration (or $\kappa$ for black holes), the Bekenstein–Hawking entropy $S=\eta A$ with $\eta=(4\hbar G)^{-1}$, and the matter flux through $\mathcal{H}$,
\begin{align}
q=-\int_{\mathcal{H}} T_{ab}\,\chi^a\,d\Sigma^b,
\end{align}
where $\chi^a$ is the approximate boost Killing vector. Demanding the Clausius relation $\delta Q=T_U\,dS$ for all local horizons yields Einstein’s equations as an equation of state. This construction is valid only in small neighborhoods where curvature, shear, and expansion can be neglected.

Jacobson’s derivation proceeds by applying the Clausius relation $\delta Q = T\, dS$ to local Rindler horizons. With the Unruh temperature $T_U=\hbar a/2\pi$, the Bekenstein–Hawking entropy $S=\eta A$, and the horizon area change from the Raychaudhuri equation,
\begin{equation}
\delta A \simeq -\int_{\mathcal{H}} \xi\, R_{ab} k^a k^b \, d\xi dA,
\end{equation}
the matter flux
\begin{equation}
q = -\kappa \int_{\mathcal{H}} T_{ab}\, \xi\, k^a k^b \, d\xi dA
\end{equation}
yields, upon imposing local equilibrium,
\begin{equation}
R_{ab} - \tfrac{1}{2} R g_{ab} + \phi g_{ab} = 8\pi G T_{ab}.
\end{equation}
Einstein’s equations thus arise as an equation of state of spacetime, with the cosmological constant emerging as an integration constant \cite{jacobson1995,padmanabhan2010}. This perspective frames gravity as emergent thermodynamics rather than a fundamental interaction, much as fluid mechanics emerges from molecular dynamics, and suggests that quantizing Einstein’s equations is no more fundamental than quantizing the ideal gas law.

\section{Modified Unruh Thermodynamics} \label{sec:mod}
Jacobson's derivation assumes the accelerated observer interacts with an infinite heat bath of infinite heat capacity. However, this assumption is physically problematic. The local Rindler horizon has finite area, limiting the available degrees of freedom. Realistic observers have finite proper lifetimes $\tau$, accessing only diamond-shaped regions rather than eternal Rindler wedges. Such observers would perceive a modified temperature that differs from the Unruh temperature when $a\tau \sim 1$. This temperature modification signals finite information content scaling with area $A$ rather than volume $V$. Consequently, the heat capacity scales as $C \sim A \sim \tau^2$. Furthermore, the finite heat bath exhibits a discrete energy spectrum and exchanges energy in quantized jumps. Thus, we identify the following reasons to consider finite heat capacity corrections to Unruh thermodynamics; (1) the quantum gravitational microstructure of spacetime implies the horizon possesses finite degrees of freedom scaling with area $N \sim A/\ell_P^2$, fundamentally bounding the heat capacity $C \sim N$. (2) realistic observers have finite lifetimes, $\tau$, and access only finite causal diamonds with bounded area $A_{\text{max}} \sim \tau^2$, making infinite heat capacity causally inaccessible. (3) thermodynamic consistency requires finite $C > 0$ to ensure physical sub-exponential growth of the density of states and allow temperature changes during energy exchange. These considerations restrict the available degrees of freedom in the thermal bath. 
A careful analysis with finite heat capacity reveals a fundamental connection to Rényi entropy, which becomes the correct entropy for the Clausius relation.

\paragraph{Finite Heat Capacity.}  Realistic local horizons of finite area $A$ must have finite heat capacity $C\sim A$. For such systems, the absorption of energy flux $q$ modifies the temperature as
\begin{equation}
T^{\text{mod}} = T_U + \frac{q}{2C}, \label{eq_teff}
\end{equation}
where the factor $1/2$ accounts for the averaging of the modified temperature during absorption, in fact, for a system with finite heat capacity $C$ that absorbs heat energy $q$, the temperature variation follows the equation $\displaystyle dT=q/C$.
 Absorption of a flux causes a linear increase in temperature from $T_U$ to $T_U+q/C$. 
The effective temperature for the Clausius relation is the average temperature during the process, so $\displaystyle T_\text{mod}=T_U+\frac{q}{2C}$. This is the standard result for thermodynamic reservoirs with finite capacities and leads to a modified entropy from the Clausius relation $q=T_{\text{mod}} S_{\text{mod}}$,
\begin{equation}
S_{\text{mod}} \simeq \frac{q}{T_U}\left(1-\frac{q}{2CT_U}\right). \label{eq_9entrp}
\end{equation}
Thus, finite-$C$ thermodynamics inevitably modifies the entropy–area law.

\paragraph{Rényi entropy from finite-$C$.}  
Expanding Rényi entropy
\begin{equation}
S_R = \frac{1}{\lambda}\ln(1+\lambda S_{BH}), \label{eq:SR}
\end{equation}
for small $\lambda$ yields
\begin{equation}
S_R = S_{BH}-\tfrac{\displaystyle\lambda}{\displaystyle2}S_{BH}^2+O(\lambda^2). \label{eq:49dd}
\end{equation}
Comparing with Eq.\eqref{eq_9entrp} identifies
\begin{equation}
\lambda = C^{-1}, \qquad \displaystyle\eta \delta A = \tfrac{\displaystyle q}{T_U}, \label{eq:lambdaC}
\end{equation}
showing that Rényi entropy naturally emerges from finite heat capacity.

\paragraph{Einstein entropy.}  
To preserve Einstein’s equations for all $C$, we introduce the \emph{Einstein entropy}
\begin{equation}
S_E = \frac{S_{BH}}{1+S_{BH}/(2C)}, \label{eq:SE}
\end{equation}
which reduces to $S_{BH}$ as $C\to\infty$. Expanding,
\begin{equation}
S_E = S_{BH}-\frac{S_{BH}^2}{2C}+O(C^{-2}), \label{eq:sEexpand}
\end{equation}
shows that $S_E$ coincides with $S_R$ at large $C$, but unlike $S_R$ it preserves the field equations exactly without corrections at any order.

\paragraph{Modified Unruh temperature.}  
With $S_{BH}=\eta A$, the Einstein and Rényi entropies read
\begin{equation}
S_E=\frac{\eta A}{1+\eta A/(2C)}, \qquad S_R=\tfrac{1}{\lambda}\ln(1+\lambda \eta A).
\end{equation}
The modified temperature Eq.\eqref{eq_teff} becomes
\begin{equation}
T_U^{mod} = T_U(1+\lambda s_{BH}), \qquad \lambda=C^{-1}, \label{eq:TRfinal}
\end{equation}
which is precisely the Rényi temperature $T_R$. Therefore, finite heat capacity leads to a consistent modification of horizon thermodynamics, with Rényi entropy emerging as the large-$C$ approximation and Einstein entropy providing the exact entropy that preserves Einstein’s equations. We point out that modifications to the entropy in the Clausius relation can stem from introducing other physical properties of the local horizon such as the local bulk viscosity of spacetime\cite{eling2006}.

\section{Modified Derivation of Einstein Equations}
\subsection{Einstein entropy: exact preservation}\label{sec:III-A}
Using the Einstein entropy $S_E$ defined in Eq.\eqref{eq:SE}, the Clausius relation becomes
\begin{equation}
q = T_U^{mod} s_E 
= T_U\!\left(1+\frac{\eta \delta A}{2C}\right)\frac{\eta \delta A}{1+\eta\delta A/(2C)}.
\end{equation}
This simplifies to the standard form
\begin{equation}
q=\frac{\hbar \kappa}{2\pi}\eta\,\delta A,
\end{equation}
showing that Einstein entropy reproduces exactly Jacobson’s derivation and thus preserves
\begin{equation}
R_{ab}-\tfrac{1}{2}Rg_{ab}+\phi g_{ab}=8\pi G T_{ab}. \label{eq:Einstein_exact}
\end{equation}
The modified Unruh temperature reads
\begin{equation}
T_U^{mod}=\frac{\hbar \kappa}{2\pi}\left(1+\frac{A}{4G\hbar C}\right).
\end{equation}
Hence, Einstein entropy is the \emph{exact entropy of local horizons}, valid for finite heat capacity.

\subsection{Rényi entropy: corrections to  GR}
Adopting instead the Rényi entropy $S_R$, the Clausius relation yields an infinite series in $\lambda=1/C$,
\begin{equation}
q = \frac{\hbar \kappa}{2\pi}\eta \left(R_{ab}k^ak^b\right)\mathcal{A}_\mathcal{H} 
+ \sum_{n\geq2} \alpha_n \left(R_{ab}k^ak^b\right)^n \mathcal{A}_\mathcal{H}^{\,n}, \label{eq:series}
\end{equation}
where $\displaystyle \alpha_n=\frac{\hbar \kappa}{2\pi}\frac{\eta^n\lambda^{n-1}}{n(n-1)}$ and $\mathcal{A}_\mathcal{H}$ is the local horizon total area. Replacing the heat flux, $q= -\kappa\ T_{ab} \,k^a k^b \mathcal{A}_\mathcal{H}$,  and simplifying we obtain, to first order in $\lambda$,
\begin{equation}\label{eq:quadraticEq}
\frac{\mathcal{A}_\mathcal{H}\lambda}{64\pi\hbar\,G^2}\mathcal{R}^2 -\frac{1}{8\pi G}\mathcal{R}+\mathcal{T}=0,
\end{equation}
where $\mathcal{R}=R_{ab}k^ak^b$ and $\mathcal{T}=T_{ab}k^ak^b$. The negative branch solution, $\mathcal{R^-}$, to the quadratic equation \eqref{eq:quadraticEq} recovers Einstein's field equation in the $\lambda\to0$ limit. Both branches read,

\begin{align}
\mathcal{R^-} &= 8 \pi  G \mathcal{T}+\frac{8 \pi ^2 \lambda\mathcal{A}_\mathcal{H}G \mathcal{T}^2}{\hbar}+O\left(\lambda ^2\right)\\
\mathcal{R^+} &= -8 \pi  G \mathcal{T}+\frac{8 G \hbar}{\mathcal{A}_\mathcal{H}\lambda }-\frac{8\pi ^2 \lambda \mathcal{A}_\mathcal{H}G \mathcal{T}^2}{\hbar}+O\left(\lambda ^2\right).
\end{align}

The branch $\mathcal{R}^-$ smoothly reduces to the Einstein relation $\mathcal{R}=8\pi G\,\mathcal{T}$ in the limit $\lambda\to0$ and therefore defines the perturbatively connected extension of general relativity. By contrast, the branch $\mathcal{R}^+$ contains a leading term with the opposite sign relative to the Einstein coupling and develops a divergence as $\lambda\to0$, indicating that it does not admit a regular Einsteinian limit. We therefore restrict attention to the branch $\mathcal{R}^-$, which captures the physically relevant finite-capacity corrections.

The ratio $\displaystyle\frac{\mathcal{A}_\mathcal{H}}{\hbar}$, can be written in term of \textit{\textit{Planck} area $l_p^2$} such as,  $\displaystyle\frac{\mathcal{A}_\mathcal{H}}{\hbar}=\displaystyle\frac{N\,l_p^2}{\hbar}=G\,N$, where $N$ is a positive integer, leading to a corrected scalar field equation,
\begin{equation}\label{eq_80fin}
    \mathcal{R} = 8 \pi  G \,\mathcal{T}+8 \pi^2\,G^2\,\lambda\, N\,\,\mathcal{T}^2+ O(\lambda^2).
\end{equation}
Introducing a universal quantity, $\nu=C/N$, as \textit{the specific heat capacity} or \textit{the heat capacity per Planck area quantum}, we get,
\begin{equation}
\mathcal{R} = 8\pi G \,\mathcal{T} 
+ \frac{8\pi^2 G^2}{\nu}\,\mathcal{T}^2 + O(\nu^{-2}),
\label{eq:correctedEinstein}
\end{equation}
A consistency bound emerges from \eqref{eq:quadraticEq},
\begin{equation}\label{eq:condition_validity}
\mathcal{T}\leq \frac{\nu}{4\pi G}.
\end{equation}
This condition puts a bound on the maximum energy flux through the local horizon. When it fails one has to add more terms from the expansion Eq.\eqref{eq:series}. 

Equation \eqref{eq:correctedEinstein} is our main result: 
\emph{finite horizon heat capacity induces quadratic corrections to Einstein’s equations on null hypersurfaces}. The Einstein entropy corresponds to the exact Einstein field equation-preserving entropy, while Rényi entropy provides a controlled expansion capturing higher-order effects.

\subsection{Numerical estimates of the energy flux density}

To assess the physical relevance of finite--heat--capacity corrections, we estimate the magnitude of the horizon energy flux density $\mathcal{T}$ in representative situations. For a unit horizon area and a characteristic proper time interval $\Delta\tau\sim \kappa^{-1}$ (equivalently $\Delta\xi\sim\kappa^{-1}$), the flux takes the parametric form
\begin{equation}
\mathcal{T}\sim \frac{\kappa^{2}}{8\pi G},
\label{eq:heat_flux}
\end{equation}
where $\kappa=a$ is the proper acceleration defining the local Rindler horizon.

Table\ref{tab:energy_flux} lists order--of--magnitude estimates of
$\mathcal{T}$, expressed in Planck units ($c=\hbar=G=1$), for a range of
physical regimes.

\begin{table}[!h]
\centering
\caption{\small Order--of--magnitude estimates of the energy flux density
$\mathcal{T}$ in Planck units ($c=\hbar=G=1$).}
\label{tab:energy_flux}
\begin{tabular}{lcc}
\toprule
\textbf{Physical regime} & \textbf{Acceleration $a$} (\si{\meter\per\second\squared}) & $\boldsymbol{\mathcal{T}}$ ($GeV^2$)\\
\midrule
Laboratory (MEMS/NEMS)\footnote{Micro/Nano-Electro-Mechanical Systems.} & $10^{8}$  & $10^{-82}$ \\
Neutron star surface & $10^{12}$ & $10^{-74}$ \\
Stellar black hole ($3M_{\odot}$) & $10^{14}$ & $10^{-70}$ \\
Laser plasma wakefield & $10^{23}$ & $10^{-52}$ \\
Heavy-ion collisions & $10^{32}$ & $10^{-34}$ \\
Planck scale & $10^{51}$ & $10^{1}$ \\
\bottomrule
\end{tabular}
\end{table}

The values show that $\mathcal{T}$ remains exceedingly small, $\mathcal{T}\ll1$, for all laboratory, astrophysical, and cosmological systems, becoming of order unity only for accelerations approaching the
Planck scale.

As a consequence, the standard Clausius relation and the associated scalar
Einstein equation,
$\mathcal{R}=8\pi\mathcal{T}$,
provide an excellent approximation across essentially all observable regimes. Finite--capacity effects enter through the corrected scalar relation
\begin{equation}
\mathcal{R}
=8\pi\,\mathcal{T}
+\frac{8\pi^{2}}{\nu}\,\mathcal{T}^{2}
+O(\nu^{-2}),
\end{equation}
where $\nu=C/N$ denotes the specific heat per Planck-area element of the local horizon. The quadratic term becomes comparable to the leading Einstein contribution only when $\mathcal{T}\sim\nu$. For $\nu$ of order unity, this occurs exclusively when $\mathcal{T}\sim1$, corresponding to Planck-scale accelerations.

These estimates indicate that the thermodynamic derivation of Einstein’s equation is extraordinarily robust, while its finite--capacity extension is relevant only in regimes where the horizon energy flux approaches the microscopic heat capacity of spacetime degrees of freedom.

\subsection{Tensorial completion and consistency}
The modified thermodynamic relation \eqref{eq:correctedEinstein} yields a scalar constraint on each local null horizon. As in Jacobson’s original construction, this relation is fundamentally
scalar and horizon-based, and does not, by itself, define a new metric theory of gravity. Nevertheless, it is legitimate to ask whether such a scalar equation admits a covariant tensorial completion consistent with standard conservation laws. 
To this end, we introduce an auxiliary null vector field $n^\mu$, normalized with respect to $k^\mu$ as
$k\!\cdot\! n=-1$.
This choice fixes the affine parametrization of the null congruence generated by $k^\mu$.
From $n^\mu$ we construct the symmetric tensor
\begin{equation}
L_{\mu\nu}=n_\mu n_\nu ,
\end{equation}
which satisfies $L_{\mu\nu}k^\mu k^\nu=1$.
Multiplying the quadratic term in Eq.\eqref{eq:correctedEinstein} by this identity and rearranging, the scalar equation may be rewritten as
\begin{equation}
\Bigl[
\bigl(R_{\mu\nu}-8\pi G T_{\mu\nu}\bigr)L_{\lambda\rho}
-\frac{8\pi^2 G^2}{\nu}\,T_{\mu\nu}T_{\lambda\rho}
\Bigr]
k^\mu k^\nu k^\lambda k^\rho=0 .
\label{eq:quartic_form}
\end{equation}
Since $k^\mu k^\nu k^\lambda k^\rho$ is totally symmetric, Eq.\eqref{eq:quartic_form} implies that the fully symmetrized tensor within brackets must vanish up to terms proportional to the metric, which do not contribute upon contraction with null vectors.
A consistent covariant completion is therefore
\begin{align}
&\bigl(R_{(\mu\nu|}-8\pi G\,T_{(\mu\nu|}\bigr)L_{|\lambda\rho)}
-\frac{8\pi^2 G^2}{\nu}\,T_{(\mu\nu}T_{\lambda\rho)} \nonumber\\
&\qquad
+ g_{(\mu\nu}M_{\lambda\rho)}
+ \phi(x)\,g_{(\mu\nu}g_{\lambda\rho)}=0 ,
\label{eq:final_tensor_eq}
\end{align}
where $M_{\mu\nu}$ is an arbitrary symmetric tensor and $\phi(x)$ is an arbitrary scalar function.
The metric terms reflect the intrinsic ambiguity of extending a null-projected scalar relation to a full tensor equation and are analogous to the emergence of the cosmological constant as an integration constant in thermodynamic derivations of gravity.
 Equation\eqref{eq:final_tensor_eq} should be interpreted as an \emph{effective consistency condition}, not as a fundamental modification of the gravitational field equations.
The quadratic term proportional to $T_{\mu\nu}T_{\lambda\rho}$ represents an effective matter self-interaction induced by finite-capacity horizon thermodynamics, rather than a modification of the geometric sector.
By construction, the scalar projection of Eq.\eqref{eq:final_tensor_eq} reproduces Eq.\eqref{eq:correctedEinstein}, while standard energy-momentum conservation, $\nabla^\mu T_{\mu\nu}=0$, remains intact.
 Geometrically, the appearance of $L_{\mu\nu}=n_\mu n_\nu$ is natural within the double-null formulation of general relativity\cite{Nakonieczna:2019,Mars:2023}, where a pair of null vectors $(k^\mu,n^\mu)$ characterizes the local structure of null hypersurfaces.
Thermodynamically, the contraction $T_{\mu\nu}n^\mu n^\nu$ corresponds to transverse stresses acting on the horizon and complements the usual energy flux $T_{\mu\nu}k^\mu k^\nu$ that enters Jacobson’s original argument.

\subsection{Compatibility with conservation laws}
A natural question concerns the compatibility of the covariant completion
Eq.\eqref{eq:final_tensor_eq} with the contracted Bianchi identities and
energy--momentum conservation.
As emphasized throughout, Eq.\eqref{eq:final_tensor_eq} is not proposed as a
fundamental modification of the Einstein equations, but as an effective
tensorial completion of a scalar, horizon-based thermodynamic relation.
Nevertheless, it is pivotal to examine its consistency at the level of
covariant conservation.

Contracting Eq.\eqref{eq:final_tensor_eq} with the inverse metric
$g^{\lambda\rho}$ yields an effective rank--two equation,
\begin{align}
&R_{(\mu}{}^{\rho} n_{\nu)} n_{\rho}
- 8\pi G\, T_{(\mu}{}^{\rho} n_{\nu)} n_{\rho}
- \frac{8\pi^2 G^2}{\nu}
\left( T_{\mu\nu} T + 2 T_{(\mu}{}^{\rho} T_{\nu)\rho} \right)
\nonumber\\
&\qquad
+ M_{\mu\nu}
+ \frac{1}{2} g_{\mu\nu} M
+ 2 \phi(x) g_{\mu\nu} = 0 ,
\label{eq:effective_rank2}
\end{align}
where $M=g^{\mu\nu}M_{\mu\nu}$.
Taking the covariant divergence and using the contracted Bianchi identity, $\nabla^\mu R_{\mu\nu}=\frac{1}{2}\nabla_\nu R$, together with $\nabla^\mu T_{\mu\nu}=0$, one obtains a consistency condition relating derivatives of the auxiliary null field $n^\mu$, quadratic matter terms, and the gauge functions $M_{\mu\nu}$ and $\phi(x)$.

Three general observations follow.

First, terms quadratic in the energy--momentum tensor, proportional to
$\nu^{-1}$, do not vanish identically under $\nabla^\mu T_{\mu\nu}=0$.
Their presence reflects effective matter self-interactions induced by finite
horizon heat capacity.
In the regime of interest, $\lambda\sim\nu^{-1}\ll1$, these contributions are
parametrically suppressed, ensuring consistency with standard low-energy
gravitational dynamics.

Second, the consistency condition couples derivatives of the auxiliary null vector $n^\mu$ to curvature and matter fields.
This indicates that $n^\mu$ cannot be chosen arbitrarily if the tensorial
completion is taken beyond its scalar projection.
This feature mirrors similar constructions in double-null and aether-like
formulations, and does not affect the validity of the underlying scalar
thermodynamic relation.

Third, the tensors $M_{\mu\nu}$ and $\phi(x)$ encode the intrinsic ambiguity
in extending a null-projected scalar equation to a fully covariant tensorial form. Their derivatives may be chosen so as to absorb residual inconsistencies.

We emphasize that these issues arise only at the level of the covariant 
completion. The fundamental result of this work—the modified scalar equation on local
Rindler horizons—remains fully compatible with conservation laws.
In particular, the Einstein entropy formulation introduced in
Sec.\ref{sec:III-A} preserves the standard Einstein equations exactly and is
therefore automatically consistent with the Bianchi identities.

\subsection{Relation to energy--momentum squared gravity}

The modified scalar relation\eqref{eq:correctedEinstein}, in which the null
projection of the Ricci tensor acquires a quadratic correction in the energy
flux $\mathcal{T}=T_{\mu\nu}k^\mu k^\nu$, bears a formal resemblance to
Energy--Momentum Squared Gravity (EMSG) models\cite{Katirci:2013okf,Roshan:2016mbt},
where Einstein’s equations are supplemented by nonlinear functions of
$T_{\mu\nu}T^{\mu\nu}$.
Our result provides an independent and complementary perspective: the quadratic
matter coupling arises here from finite--heat--capacity horizon thermodynamics,
rather than from a modification of the gravitational action.

A key distinction concerns both interpretation and structure.
In EMSG, quadratic terms originate from variations of $T_{\mu\nu}T^{\mu\nu}$
and typically lead to effective violations of covariant energy--momentum
conservation, $\nabla_\mu T^{\mu\nu}\neq0$, unless additional constraints are
imposed.
By contrast, in the present framework the fundamental relation is a scalar,
horizon-based thermodynamic equation, and standard conservation laws are
preserved at the level of the underlying Einstein equations.

Moreover, the null projection highlights a structural difference.
For perfect-fluid matter, the EMSG correction projected along a null direction
generically produces contributions linear in $\mathcal{T}$, whereas finite
heat-capacity effects derived here yield corrections quadratic in
$\mathcal{T}$.
This difference reflects the distinct physical origins of the two approaches:
EMSG encodes modified matter couplings at the level of the action, while the
present corrections arise as effective matter self-interactions induced by
finite-capacity horizon thermodynamics.

Despite these differences, the comparison establishes a nontrivial bridge
between horizon thermodynamics and effective gravity models with higher-order
matter couplings, suggesting that EMSG-like structures may admit a natural
thermodynamic interpretation.

\section{Conclusion}

Jacobson’s insight that Einstein’s equations emerge as an equation of state
recasts general relativity as an intrinsically thermodynamic description of
spacetime. Building on this paradigm, we have shown that incorporating finite
heat-capacity effects into Unruh thermodynamics leads to well-defined
subleading corrections while preserving the Einstein equations at leading
order. The resulting connection to Rényi entropy naturally embeds emergent
gravity within a quantum-information–theoretic framework, where horizon
thermodynamics and entanglement equilibrium play a central role.

The modified thermodynamic relation derived here remains fundamentally scalar
and horizon-based, as in Jacobson’s original construction. Its covariant
completion should therefore be understood not as defining a new fundamental
metric theory, but as a consistency framework illustrating how finite-capacity
corrections may be incorporated without violating standard conservation laws.
In this sense, the associated tensorial structures encode effective matter
self-interactions induced by horizon thermodynamics rather than genuine
modifications of the gravitational sector.

Remarkably, the resulting quadratic dependence on the null energy flux echoes
features of energy–momentum squared gravity models, suggesting that higher-order matter couplings may admit a thermodynamic origin. More broadly, the emergence of an undetermined scalar function multiplying the metric reflects a structural freedom familiar from thermodynamic and Weyl-invariant formulations of gravity, where such terms arise as integration data rather than fixed parameters.

These results sharpen the thermodynamic foundations of emergent gravity and
open new avenues for exploring finite-capacity effects in both theoretical and phenomenological settings, ranging from analog gravity systems to strongly coupled quantum matter.

\section*{Acknowledgements}
\paragraph{}H. El M would like to acknowledge the networking
support of the COST Action 
 CA 22113 - Fundamental challenges in theoretical physics (Theory and Challenges), 
CA 21136 - Addressing observational tensions in cosmology with systematics and fundamental physics (CosmoVerse), and 
CA 23130 - Bridging high and low energies in search of quantum gravity (BridgeQG). He also thanks IOP for its support. This work was carried out under the project UIZ 2025 Scientific Research Projects: {\tt PRJ-2025-81}.

\bibliographystyle{unsrt}
\bibliography{references}

\end{document}